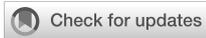

# High-resolution synthesis of high-density breast mammograms: Application to improved fairness in deep learning based mass detection

Lidia Garrucho[1*], Kaisar Kushibar[1], Richard Osuala[1], Oliver Diaz[1], Alessandro Catanese[2], Javier del Riego[3], Maciej Bobowicz[4], Fredrik Strand[5], Laura Igual[1] and Karim Lekadir[1]

[1]Barcelona Artificial Intelligence in Medicine Lab, Facultat de Matemàtques i Informàtica, Universitat de Barcelona, Barcelona, Spain, [2]Unitat de Diagnòstic per la Imatge de la Mama (UDIM), Hospital Germans Trias i Pujol, Badalona, Spain, [3]Área de Radiología Mamaria y Ginecólogica (UDIAT CD), Parc Taulí Hospital Universitari, Sabadell, Spain, [4]2nd Department of Radiology, Medical University of Gdansk, Gdansk, Poland, [5]Breast Radiology, Karolinska University Hospital and Department of Oncology-Pathology, Karolinska Institutet, Stockholm, Sweden

Computer-aided detection systems based on deep learning have shown good performance in breast cancer detection. However, high-density breasts show poorer detection performance since dense tissues can mask or even simulate masses. Therefore, the sensitivity of mammography for breast cancer detection can be reduced by more than 20% in dense breasts. Additionally, extremely dense cases reported an increased risk of cancer compared to low-density breasts. This study aims to improve the mass detection performance in high-density breasts using synthetic high-density full-field digital mammograms (FFDM) as data augmentation during breast mass detection model training. To this end, a total of five cycle-consistent GAN (CycleGAN) models using three FFDM datasets were trained for low-to-high-density image translation in high-resolution mammograms. The training images were split by breast density BI-RADS categories, being *BI-RADS A* almost entirely fatty and *BI-RADS D* extremely dense breasts. Our results showed that the proposed data augmentation technique improved the sensitivity and precision of mass detection in models trained with small datasets and improved the domain generalization of the models trained with large databases. In addition, the clinical realism of the synthetic images was evaluated in a reader study involving two expert radiologists and one surgical oncologist.

KEYWORDS

data synthesis, full-field digital mammograms, generative adversarial networks (GANs), data augmentation (DA), mass detection, reader study, breast cancer

# 1 Introduction

Breast density is divided into four categories in the American College of Radiology Breast Imaging and Data System (ACR BI-RADS) 5$^{th}$ edition (1). The categories range from A to D and correspond to fatty, scattered, heterogeneous, and extremely dense breasts. The qualitative classification of breast density in mammography is an accepted method in breast radiology with good inter-observer and intra-observer agreement (2), despite the fact that commercial software can produce a more accurate quantitative measure by calculating the ratio of fibroglandular tissue to the total breast area.

In mammography databases, the distribution of breast densities among women aged 40 years or older is approximately 43% for dense breasts: 36% for BI-RADS C and 7% for BI-RADS D (3). Han et al. (4) found that women with a family history of breast cancer were more likely to have dense breasts than women with no cancer in the family history. In addition, high breast density is associated with an increased risk of interval cancers (5), being those 13–31 times more likely in BI-RADS D breasts than in BI-RADS A (6–8). Consequently, it is recommended to decrease the interval between screening mammograms and consider supplemental screening for women with dense breasts (8, 9).

Dense breast tissue is one of the strongest and most common independent risk factors for breast cancer (5, 6, 10). On mammograms, masses and other suspicious findings can be obscured in normal dense tissue and become imperceptible on mammograms. Therefore, the sensitivity of mammography decreases with increasing breast density and has a range value of 81-93% for fatty breasts, 84-90% for breasts with scattered fibroglandular density, 69-81% for heterogeneously dense breasts, and 57-71% for extremely dense breasts in women 40-74 years of age (5). Although mammography is the gold standard non-invasive method for breast cancer detection in population-based screening, women with dense breasts have shown both a reduced cancer detection and higher mortality rates (11–13).

The goal of this study is two-fold. First, mitigate the differences in computer-aided detection (CADe) systems sensitivity by breast density (14). Second, improve the performance of state-of-the-art deep learning-based breast mass detection models by the means of synthetic data augmentation.

Figure 1 shows the differences in sensitivity of our deep learning-based mass detection model by density composition. The breast composition distribution of the dataset used to train the model has a big unbalance in categories A (9%) and D (5%) (Figure 1A). Nonetheless, the sensitivity between fatty (95% for BI-RADS A) and extremely dense breasts (82% for BI-RADS D) is very different (Figure 1B). The decrease in performance is partly caused by the high rate of false positives in extremely dense breasts (Figure 1C).

Data augmentation is commonly used to increase the variability of the training samples, improve the model generalization and avoid overfitting. Among all deep learning-based augmentation techniques, Generative Adversarial Networks (GANs) (15) are frequently used to generate new synthetic samples in an unsupervised manner. GANs have been previously used to synthesize full-field digital mammograms (FFDMs) or lesion patches, normally at low resolutions (16–18). Becker et al. (19) trained a cycle-consistent GAN (CycleGAN) on downscaled mammograms (256×256 and 515×408 pixels) to artificially inject or remove suspicious features. In their reader study, three radiologists could discriminate between original and synthetic images with an area under the curve (AUC) of 0.94, mainly due to the presence of artifacts. Zakka et al. (20) trained a style-based GAN to generate 512×512 mammograms enabling user-controlled global and local attribute-editing. Then, a double-blind study involving four expert radiologists assessed the quality of the resulting images achieving an average AUC of 0.54.

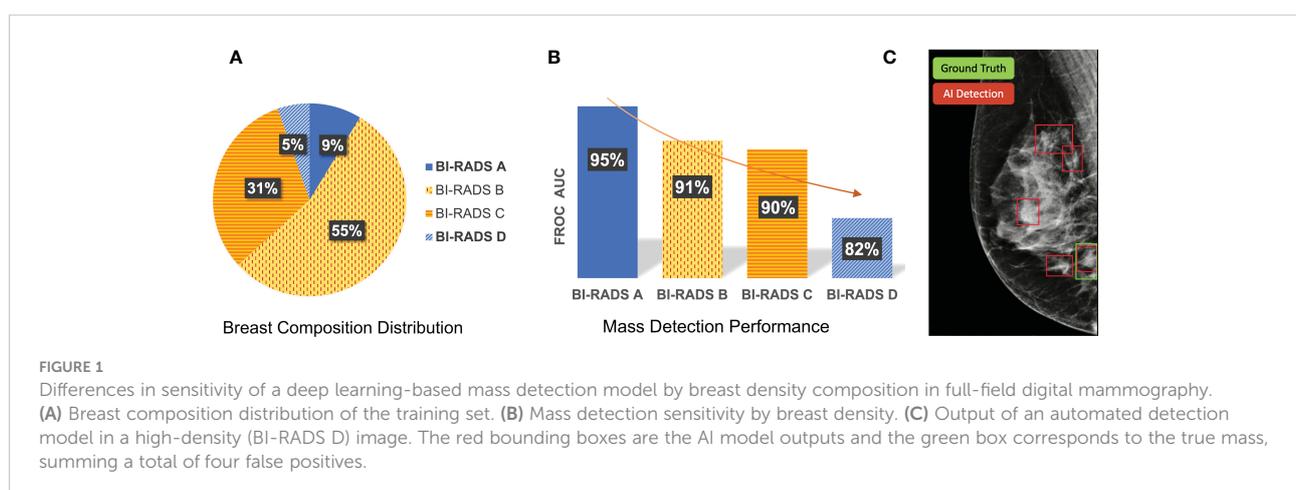

FIGURE 1
Differences in sensitivity of a deep learning-based mass detection model by breast density composition in full-field digital mammography.
**(A)** Breast composition distribution of the training set. **(B)** Mass detection sensitivity by breast density. **(C)** Output of an automated detection model in a high-density (BI-RADS D) image. The red bounding boxes are the AI model outputs and the green box corresponds to the true mass, summing a total of four false positives.

Other studies used the synthetic mammograms as data augmentation to improve the performance in different downstream tasks. Synthetic data augmentation using GANs was evaluated in breast cancer classification by Shrinivas et al. (21). The proposed model, a Deep Convolutional GAN (DCGAN), synthesized FFDM with 256×256 image resolution. The synthetic images were validated using a Visual Turing test with the help of medical experts and were easily spotted by the radiologist because they lacked the sharpness and fine-grained details of original mammograms. In a similar study, Jendele et al. (22) balanced the ratio of benign and malignant lesions in the training set using a CycleGAN trained to translate healthy mammograms to mammograms containing malignant findings. The synthetic mammograms were 256×256 pixels, as higher image resolutions introduced many artifacts. The benefits of using the synthesized mammograms for data augmentation were inconclusive considering that the performance of their detection model did not improve. Muramatsu et al. (23) trained a CycleGAN using masses from unrelated domains – lung CT and mammography – to synthesize 256×256 pixels masses and improve the mass classification in mammography. However, no statistical difference was found between the model trained with synthetic masses and the classifier trained with original mass patches.

In this study, the original resolution of FFDMs, around 5 Megapixels (MP), with image sizes of 3328×4084 or 2560×3328 pixels depending on the manufacturer. Two main challenges have prevented the use of GANs for high-resolution FFDM synthesis. The first one is the high demand for graphics processing unit (GPU) memory, which typically scales with the input and output resolutions. As an example, CycleGAN needs more than 24GB of GPU memory when the input image is larger than 1MP. The second challenge is data scarcity. The training set has to be representative enough to generate realistic samples and overcome the training instabilities and image artifacts of GANs. High intrinsic heterogeneity exists across mammograms due to the huge variability of breast sizes, shapes, and compositions. Moreover, FFDMs contain very fine structural details at high resolution such as the different parenchymal patterns, nipples, and pectoral muscles, the presence of lymph nodules, microcalcifications, or calcifications, among many other associated features.

Korkinof et al. (24) was the first study that managed to synthesize high-resolution FFDM images using a Progressively Growing GAN (PGGAN) (25). Their PGGAN was trained using more than 400,000 FFDMs and was demonstrated to generate mammograms up to a 1280×1024 pixel resolution. All mammograms available were used for training, independently of the clinical findings, and only images with post-operative artifacts and large foreign bodies such as implants were excluded from the training set. In a separate retrospective study, Kornikof et al. (26) evaluated the perceived realism of the synthetic FFDM images in a reader study involving 55 radiologists and 62 non-radiologists. Overall, in the setup of this study, the synthetic images were shown to be indistinguishable from original mammograms. However, it was unclear and was not further investigated whether the synthetic FFDMs have relevant applications for clinical purposes.

This work presents for the first time the use of GAN models to generate high-resolution FFDMs with increased breast density using images from different manufacturers and datasets. Moreover, we evaluated the potential of using the synthesized images as data augmentation to improve the mass detection performance. Only a single prior study had performed similar data augmentation by breast density categories for improved mass detection (27). However, the authors employed mathematical breast phantoms generated using the pipeline in the VICTRE study (28) instead of GANs. The breast phantoms were generated across the four BI-RADS breast density categories for each view, cranio-caudal (CC) and mediolateral oblique (MLO), and were modeled after a single vendor, the Siemens Mammomat Inspiration. The limitation of their study is the lack of diversity within each density type, including the size and shape of the breast. Moreover, the statistical analysis did not show a significant difference between the Free-response Receiver Operating Characteristic (FROC) curves for mass detection.

To summarize, the contributions of our study are as follows:

- Synthesize high-resolution high-density FFDMs using GAN-based models from three different datasets and mammography systems (manufacturers).
- Tackle the class imbalance by breast density composition by augmenting the training set using high-density synthetic mammograms.
- Improve the performance of mass detection in extremely dense breasts, categorized as BI-RADS D.
- Investigate the potential of high-density data augmentation for domain adaptation.
- Evaluate the anatomical realism of the synthetic mammograms in a reader study involving two expert radiologists and one surgical oncologist.

# 2 Materials and methods

## 2.1 Datasets and breast density

Four different datasets were used in this study. General details of these datasets are presented in Table 1. The OPTIMAM Mammography Image Database (OMI-DB) (29) comprises FFDM from the Breast Screening Programme of the United Kingdom (UK). In this study, we used the subset of mammograms captured with the Hologic Selenia Dimensions scanners (Hologic, Inc., Massachusetts, United States). The proportion of fibroglandular (dense) tissue in the breast was

TABLE 1 Digital mammography datasets used in this study.

| | BCDR | | CSAW | | OPTIMAM Hologic | | | INbreast | |
|---|---|---|---|---|---|---|---|---|---|
| | ACR | Normal MMG | LIBRA (%) | Normal MMG | Volpara VBD(%) | Normal MMG | MMG with masses | ACR | MMG with masses |
| BI-RADS A | 1 | 40 | ≤ 2.8 | 435 | ≤ 3.5 | 972 | 344 | 1 | 42 |
| BI-RADS B | 2 | 40 | (2.8, 25) | 52064 | (3.5, 7.5) | 3670 | 1740 | 2 | 36 |
| BI-RADS C | 3 | 62 | (25, 75) | 38545 | (7.5, 15.5) | 1987 | 808 | 3 | 21 |
| BI-RADS D | 4 | 58 | ≥ 75 | 394 | ≥ 15.5 | 708 | 161 | 4 | 8 |

For each dataset, the total number of mammograms (MMG) with the available breast density information are mapped to the corresponding BI-RADS categories. In CSAW, the breast percentage was obtained running LIBRA software. In OPTIMAM Hologic, the Volumetric Breast Density (VBD) percentage was obtained using Volpara software. In BCDR and INbreast, only the American College of Radiology (ACR) categories are available in the dataset information.

obtained from the commercial Volpara software (version 1.5.4, Volpara Health, Wellington, New Zealand). The Volumetric Breast Density (VBD) percentage of each mammogram was mapped to the corresponding BI-RADS breast density category. Only normal mammograms – without pathologies – were selected to train the synthetic data generation models. The mammograms containing masses were used to train and test the mass detection AI model.

The non-hidden case-control dataset of the CSAW dataset (30) was also used in this study. The dataset comprises screening FFDMs from Karolinska University Hospital (Solna, Sweden) acquired with Hologic Inc. devices. A total of 91,484 normal FFDM contained breast density information estimated using the LIBRA automated tool (University of Pennsylvania, United States) (31). The mapping between LIBRA breast density percentage and BI-RADS A and D categories was done by selecting the tails of the percent density distribution of the healthy exams.

The Breast Cancer Digital Repository (BCDR) dataset (32) is a public dataset from 2012 comprising images supplied by the Centro Hospitalar São João, at University of Porto (Portugal) and obtained using a MammoNovation Siemens FFDM scanner. For our purposes, we selected a total of 200 FFDMs without pathologies. The breast density categories were provided in the annotations of the dataset following the American College of Radiology (ACR) statement on reporting breast density (1).

The INbreast dataset (33) was acquired from a single Portuguese center using a FFDM system, the Siemens MammoNovation. INbreast was used in this study as an external validation dataset for the mass detection model trained with OPTIMAM Hologic images. In addition, INbreast was used to train a mass detection model in a low data regime scenario. The dataset contains 107 FFDMs with 116 annotated masses from different breast densities. The percentage of images in each BI-RADS category is 36%, 35%, 22% and 7% for BI-RADS A, B, C and D.

All images have a matrix of 3328×4084 or 2560×3328 pixels, depending on the compression plate used for image acquisition. In a previous work, we confirmed that a resolution of 1332×800 pixels was enough to detect small masses and reach state-of-the-art performance in different AI detection methods (34). All the FFDM images were cropped to the breast region and resized to 1332×800 pixels keeping the aspect ratio. Our target resolution for data synthesis was the same as the one used by our deep learning mass detection model.

## 2.2 Synthesis of high-density full-field digital mammograms

Our goal was to synthesize high-density FFDM from original low density mammograms and then use the synthetic data to improve the performance of our mass detection models. To this end, the training images were split by breast density BI-RADS categories, being *BI-RADS A* the source domain and *BI-RADS D* the target domain. Before training, all mammogram images were resized to the target resolution (1332×800 pixels), the input size of the mass detection model.

The CycleGAN (35) was the method selected to perform the low-to-high-density mammogram translation. The choice of CycleGAN was motivated by its widespread use and successes reported in the cancer imaging domain (17). A key methodological feature of CycleGAN is that its training data can be unpaired without the need for corresponding image pairs in source and target domains. Unpaired training data ensured the applicability of CycleGAN to our datasets, in which image pairs are not available, as the same breast of the same patient cannot be from both the high and the low breast density domains.

As shown in Figure 2, the CycleGAN contains two mapping functions: 1) an image $x$ in the source domain is mapped to a synthetic image $G(x) = \hat{y}$ in the target domain *via* a generator $G$; and 2) an image $y$ is mapped from target to source $F(y) = \hat{x}$ *via* a generator $F$. This enables translating images from source to target and back to the source domain $F(G(x)) = \hat{x}$. Both generators $F$ and $G$ are paired with corresponding discriminators, which try to classify whether a generated image is real or synthetic in a two-player *minmax* game with their respective generator (15). Based on the predictions of the

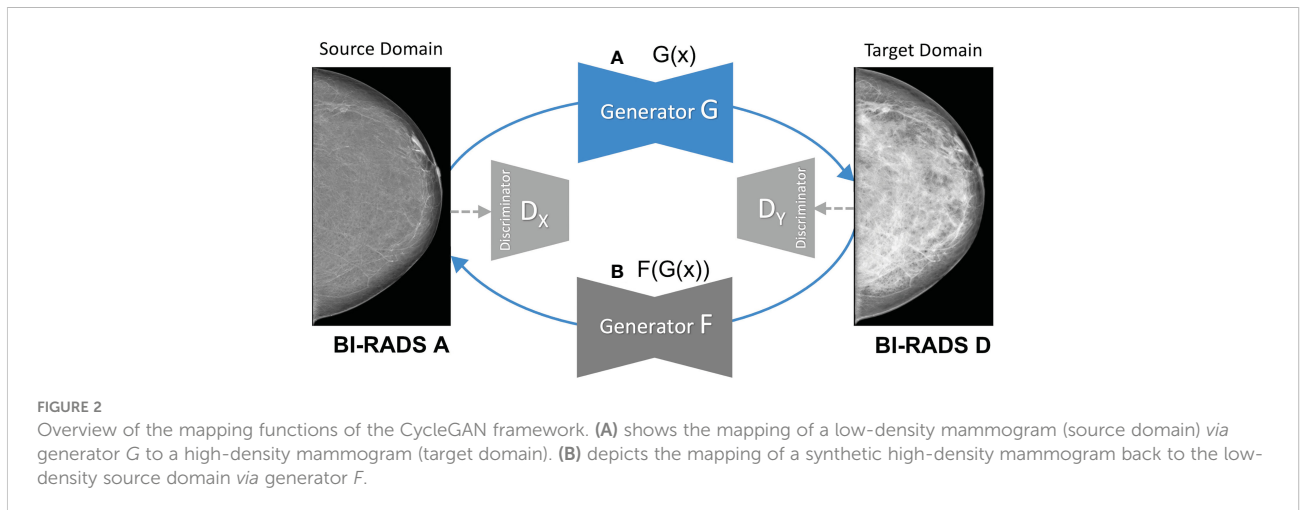

FIGURE 2

Overview of the mapping functions of the CycleGAN framework. **(A)** shows the mapping of a low-density mammogram (source domain) via generator G to a high-density mammogram (target domain). **(B)** depicts the mapping of a synthetic high-density mammogram back to the low-density source domain via generator F.

discriminator, binary cross entropy is used to compute the adversarial loss as shown below, which is back-propagated to the respective generator network.

$$\mathcal{L}_{\text{GAN}}(G, D_Y, X, Y)$$
$$= \mathbb{E}_{y \sim p_{\text{data}}(y)}[\log D_Y(y)] + \mathbb{E}_{x \sim p_{\text{data}}(x)}[\log (1 - D_Y(G(x)))]$$

CycleGAN further contains two cycle-consistency losses defined as $L1$ reconstruction loss between (i) the source image $x$ and the reconstructed source image $F(G(x)) = \hat{x}$ and between (ii) the target image $y$ and the reconstructed target image $G(F(y)) = \hat{y}$.

$$\mathcal{L}_{\text{cyc}}(G, F) = \mathbb{E}_{x \sim p_{\text{data}}(x)}[\| F(G(x)) - x \|_1]$$
$$+ \mathbb{E}_{y \sim p_{\text{data}}(y)}[\| G(F(y)) - y \|_1]$$

As defined by (35), the full loss function of our CycleGAN reads as follows with the λ parameter (λ=10) weighting the relative importance between cycle-consistency and adversarial losses.

$$\mathcal{L}(G, F, D_X, D_Y) = \mathcal{L}_{\text{GAN}}(G, D_Y, X, Y) + \mathcal{L}_{\text{GAN}}(F, D_X, Y, X) + \lambda \; \mathcal{L}_{\text{cyc}}(G, F)$$

Only healthy mammograms (Normal) were used to train the different CycleGAN models. The main reason was to avoid feature hallucinations that have been shown to occur in cancer imaging when training on images where tumors were present (17, 36).

A total of five CycleGAN models from the three different datasets –BCDR, CSAW and OPTIMAM– were trained (see Figure 3). For the OPTIMAM and the CSAW dataset, a different model was trained for each view (CC and MLO). That was because the anatomic features of CC and MLO are different and more specialized CycleGAN models – focusing only on one view – should learn better translations. However, the small sample size in BCDR dataset made it unfeasible to split the models by view and, for this dataset, a single model was trained combining both CC and MLO views (Figure 3A).

The models were trained using a single GPU (24GB NVIDIA GeForce RTX 3090) for a maximum of 200 epochs, using a batch size of 1 and adjusting the learning rate following the recommendations implemented in the CycleGAN Pytorch framework[1]. The models are available inside the mediGAN library (37).

## 2.3 Mass detection using high-density synthetic data augmentation

### 2.3.1 Effectiveness of the data augmentation strategies

Data scarcity is an important topic in the cancer imaging field and can substantially impact the performance of AI models. In this regard, the effectiveness of the proposed data augmentation technique will depend on the available training images. Ideally, the mass detection models should be trained with large databases with enough representation of the different breast density categories. However, as described in Table 1, there is a considerable imbalance among BI-RADS categories in the datasets. To evaluate the effectiveness of the proposed data augmentation under different data availability conditions, we have designed four different training scenarios.

First, we analyzed the impact of the dataset size. Both a large and a small public dataset were selected to train the detection models, respectively, in high- and low-data regime scenarios. The OPTIMAM Hologic database was selected to investigate the first scenario, comprising more than 3000 mammograms with annotated masses. The low-regime scenario was simulated using the INbreast dataset, which contains 107 FFDMs with annotated masses.

Second, we evaluated how well the synthetic images were able to simulate extremely dense breasts (BI-RADS D) during training. To this end, we trained the detection models twice. First, removing the real BI-RADS D images from the training set and keeping only real mammograms from BI-RADS A, B, and C. In this training setup, the model did not see any real BI-RADS D images during training. Second, including the real BI-RADS D mammograms in the training. In this last scenario, the OPTIMAM Hologic detection models used 25% of the real BI-RADS D mammograms available for training while the remaining ones were used for testing and validation purposes. Note that the INbreast dataset has only 8 real BI-RADS D mammograms available for training.

---

1 https://github.com/junyanz/pytorch-CycleGAN-and-pix2pix

### 2.3.2 Training and testing

Our baseline model for mass detection in FFDM is a Deformable DETR (38) with a ResNet50 (39) feature extraction backbone. The model choice was based on the good performance of Deformable DETR in our previous comparative study (34).

First, the baseline model was trained without synthetic data augmentation. Then, four mass detection models were trained with different data augmentation strategies as follows. Three mass detection models, named *BC-Aug*, *CS-Aug* and *OP-Aug*, used synthetic images from a single CycleGAN model with a proportion of 1:1 per mammogram – 1 real and 1 synthetic. The fourth detection model, named *OP-CS-BC-Aug*, included a combination of synthetic images from all the CycleGAN models with a proportion of 1:3 per mammogram – 1 real and 3 synthetic. Thus, the combined detection model was trained with a proportion of synthetic data three times higher than the other three models.

The *BC-Aug* detection model was trained using synthetic data generated from the *BC-All* CycleGAN (Figure 3). Similarly, the *OP-Aug* and the *CS-Aug* detection models were trained with synthetic data from the corresponding CycleGAN models. Note that *BC-All* generates both the MLO and CC views, whereas the CSAW and OPTIMAM have two independent models for each view: *CS-CC*, *CS-MLO*, *OP-CC* and *OP-MLO*.

Only random flipping was used as additional data augmentation. All models were trained five times, that is, using five different random seeds, and evaluated by averaging the results across seeds. A single GPU was used (24GB NVIDIA GeForce RTX 3090) to train each model for a maximum of 60 epochs, using a batch size of 1 and adjusting the learning rate following the implementation recommendations of the MMDetection framework[2].

### 2.3.3 Evaluation metrics and statistical significance tests

The area under the curve (AUC) of the Free-response Receiver Operating Characteristic (FROC) curve (40) was used to compare the baseline with the different data augmentation strategies. The AUC was computed by varying the confidence threshold of each bounding box in a range of FPPI ∈ (0, 1) (False Positives per Image). If the Intersection-over-Union (IoU) of the prediction and the ground truth was greater than 10%, then the bounding box was considered as a True Positive (TP) (41).

To assess statistical differences in AUCs between the baseline and the models trained with different data augmentation strategies, we used the paired version of DeLong's test for ROC curves (42). To do so, we defined a maximum of 10 False Positives per Image (FPPI) and compared the detection scores of the baseline with the augmented models. The statistical analysis was performed using the code from the fast implementation of DeLong by (43).

## 2.4 Reader study

A reader study involving two breast radiologists and one surgical oncologist specializing in breast disease was conducted to determine whether the synthetic images were distinguishable from the real ones as a proxy for perceptual realism. The readers were two breast radiologists from different hospitals in Spain, with +9 (Reader A) and +7 (Reader B) years of experience and the surgical oncologist from a hospital in Poland with +12 years of experience in image guided breast biopsy and lesion localization techniques (Reader C).

The reader study contained 180 high-density mammograms balanced by view and dataset. A total of 90 images were original BI-RADS D mammograms: 30 from OPTIMAM Hologic, 30 from CSAW and 30 from BCDR dataset. The other 90 images were synthetic mammograms generated with the different CycleGAN models: 30 images from *OP-CC* and *OP-MLO* models, 30 from *CS-CC* and *CS-MLO* models, and 30 from *BC-All* model. The original low-density mammograms used to generate the synthetic images were randomly selected from BI-RADS A OPTIMAM Hologic dataset.

The reader study was designed as a stand-alone ImageJ[3] plugin. A single mammogram was displayed at a time (Figure 4B) next to a multiple-choice panel (Figure 4A) to assign a label based on the certainty of the image being synthetic (Fake) or original (Real). The 6 different choices were converted to equally-distributed probabilities of (0.95, 0.77, 0.59, 0.41, 0.23, 0.05) to compute the ROC curve of each reader as in Alyafi et al. (44). No feedback was given to the readers during the assessment to avoid the identification of artifacts of synthetic images.

The images were resized to a maximum 532 pixels height to avoid the identification of the checkerboard artifacts and the lack of sharpness of synthetic mammograms, which is related to upsampling (45) in GANs. The goal of the reader test was to evaluate the anatomically-plausible realism of the synthetic images rather than the noise and common artifacts of the CycleGAN models. Additionally, we asked radiologists to identify the artifacts and common pitfalls of the synthetic images after performing the reader study (Section 4.3.2.1).

---

2 https://github.com/open-mmlab/mmdetection

3 https://imagej.nih.gov/ij/

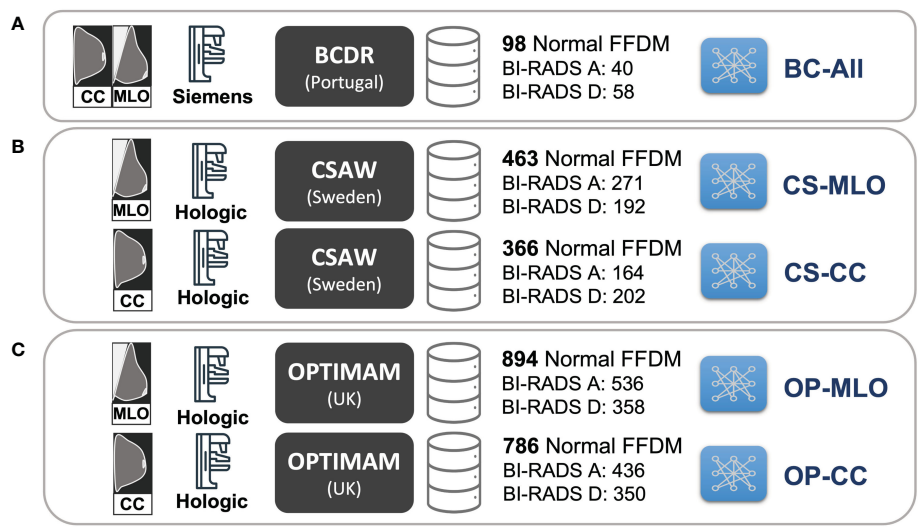

FIGURE 3
Training setup for the different CycleGAN models. **(A)** *BC-All* model for both CC and MLO views trained with 98 normal FFDMs from BCDR dataset. **(B)** two models for CC (*CS-CC*) and MLO (*CS-MLO*) views trained with 463 and 366 normal FFDMs from CSAW dataset. **(C)** two models for CC (*OP-CC*) and MLO (*OP-MLO*) views trained with 894 and 786 normal FFDMs from OPTIMAM Hologic dataset.

## 3 Results

### 3.1 Evaluation for the CycleGAN models

The Fréchet Inception Distance (FID) (46) is a useful metric to measure the quality of the synthetic mammograms and compare the synthetic models with each other. The FID was calculated between two different sets of images [4]. Since FID is not an absolute measure, we defined lower and upper bounds using real mammograms. First, the lower bound was defined as the FID between two different splits of real BI-RADS D mammograms from the same dataset. Second, the upper bound was given by the FID between real BI-RADS A and BI-RADS D mammograms. The synthetic BI-RADS D images in our evaluation set were generated from real BI-RADS A mammograms using the five different CycleGAN models (*BC-All*, *CS-CC*, *CS-MLO*, *OP-CC* and *OP-MLO*). The BI-RADS A mammograms were from the OPTIMAM Hologic and INbreast datasets, the training datasets for the mass detection models. In such manner, we would like to evaluate if the CycleGAN models were able to translate from the source domains – OPTIMAM Hologic and INbreast – to the target domains – OPTIMAM Hologic, CSAW or BCDR – with enough fidelity.

In Table 2, the FID between different groups of images is shown. Overall, the FID score was lower for the synthetic CC mammograms than for the MLO view. Ideally, the FID between synthetic and real BI-RADS D mammograms should be as close as possible to the lower bound and do not exceed the upper bound. As an example, the FID scores in CSAW CC view between real and synthetic BI-RADS D for both OPTIMAM Hologic and INbreast input images, were 99.95 and 124.84, which are between the lower bound (42.57) and the upper bound (142.34). For the synthetic BI-RADS D images generated from OPTIMAM Hologic, both in OPTIMAM Hologic and CSAW CycleGANs, the FID between the real and the synthetic BI-RADS D mammograms lies between the FID bounds. Only in BCDR, the FID score was greater than the FID between both original images.

On the contrary, for the synthetic BI-RADS D images generated from INbreast, the FID was lower in BCDR than in OPTIMAM Hologic and CSAW. Both BCDR and INbreast were acquired with Siemens scanners, while OPTIMAM and CSAW were acquired with Hologic Inc. scanners. This indicates a less pronounced domain difference between BCDR and INbreast, which aligns with the correspondingly smaller FID. Lastly, the synthetic images from the OPTIMAM Hologic CycleGAN models lie outside the bounds when the input images came from INbreast. In section 4.2, we will evaluate the impact on mass detection performance when synthetic images with high FID scores are used as data augmentation.

---

4 GitHub repository used to compute FID: https://github.com/mseitzer/pytorch-fid (commit 3d604a2)

### 3.1.1 Qualitative analysis of the synthetic images

In Figure 5 there are some sample high-density mammograms generated with the different CycleGAN models. The first row (Figure 5A) corresponds to the CC view and the second (Figure 5B) to MLO. The first column is the original BI-RADS A FFDMs from OPTIMAM Hologic scanner. The next columns are the synthetic FFDMs from the different CycleGAN models. By visual inspection, one can see that the synthetic images did not remove the masses in the original mammograms, which enabled their usage for data augmentation in the mass detection training. We can also confirm that the synthetic images are able to properly translate from source to target domain. The low-to-high-density image translation was applied for the input BI-RADS A, B, or C. In higher density mammograms, such as BI-RADS C, the changes were more subtle and less density was added to the output image.

## 3.2 Mass detection performance

All models were evaluated in independent sets of 120 mammograms of each BI-RADS category from the OPTIMAM Hologic dataset. As our objective was to improve the detection performance in BI-RADS D mammograms, we focused on the performance gain in BI-RADS D mammograms. The evaluation metrics for the other BI-RADS categories (A, B and C) can be found in the Supplementary Material.

### 3.2.1 Large data availability scenario

The corresponding FROC curves of the detection models trained with OPTIMAM Hologic are shown in Figure 6. All

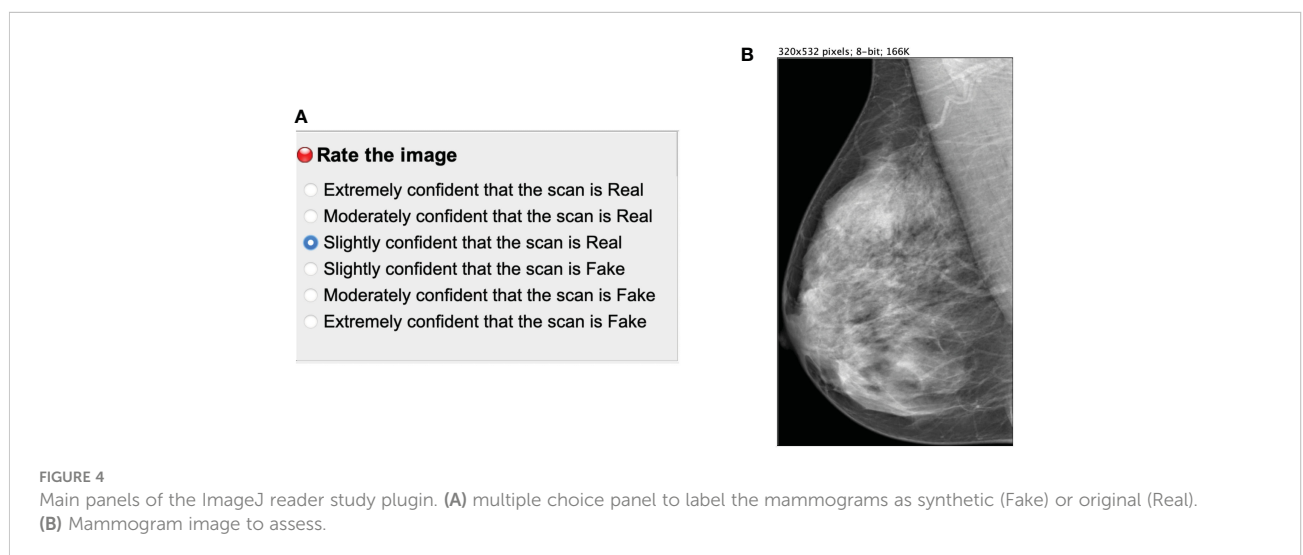

FIGURE 4
Main panels of the ImageJ reader study plugin. **(A)** multiple choice panel to label the mammograms as synthetic (Fake) or original (Real). **(B)** Mammogram image to assess.

TABLE 2 Fréchet Inception Distance between different sets of images.

|  | CycleGAN Dataset | View | Real BI-RADS D (lower bound) | Synthetic BI-RADS D (input BI-RADS A from OPTIMAM Hologic) | Synthetic BI-RADS D (input BI-RADS A from INbreast) | Real BI-RADS A (upper bound) |
|---|---|---|---|---|---|---|
| Real BI-RADS D | BCDR | CC + MLO | 66.16 | **149.10** (*BC-All*) | **103.98** (*BC-All*) | 142.61 |
|  | CSAW | CC | 42.57 | **99.95** (*CS-CC*) | **124.84** (*CS-CC*) | 142.34 |
|  |  | MLO | 73.54 | **165.48** (*CS-MLO*) | **183.89** (*CS-MLO*) | 206.04 |
|  | OPTIMAM Hologic | CC | 34.17 | **73.16** (*OP-CC*) | **132.04** (*OP-CC*) | 107.99 |
|  |  | MLO | 57.24 | **109.68** (*OP-MLO*) | **175.63** (*OP-MLO*) | 140.93 |

The lower bound was defined as the FID between two different splits of real BI-RADS D mammograms from the same CycleGAN dataset. Similarly, the upper bound was given by the FID between real BI-RADS A and BI-RADS D mammograms. The synthetic BI-RADS D images were generated from real BI-RADS A mammograms from OPTIMAM Hologic and INbreast datasets. The different CycleGAN models (*BC-All*, *CS-CC*, *CS-MLO*, *OP-CC* and *OP-MLO*) were used to generate the synthetic images. Bold values indicates the FID values from the synthetic images.

detection models were evaluated on BI-RADS D test set from OPTIMAM Hologic while the INbreast dataset was used as an external validation set.

The models trained using only synthetic BI-RADS D mammograms in training obtained more benefit from the high-density data augmentation. Table 3 summarizes the gains in AUC and the *p-values* for the best performing data augmentation strategies. When only synthetic BI-RADS D images were present in training, the combined data augmentation strategy (*OP-CS-BC-Aug*) obtained a gain of +1.24, increasing the AUC from 79.71% to 80.95% with a *p-value* of 0.0696. The *OP-CS-BC-Aug* obtained a +2.95 gain in AUC in the external validation dataset – INbreast. This confirmed that the resulting model is more robust in the presence of domain-shifts compared to the baseline. However, the best performing data augmentation strategy for the INbreast test set was the *BC-Aug*, with a total gain of +4.15 in AUC. The models trained with real and synthetic BI-RADS D mammograms obtained less gain in the OPTIMAM Hologic test set. To this end, the most substantial gain was achieved with the *BC-Aug* model, with +0.5 in AUC and a *p-value* of 0.2277, which indicates it is not statistically significant. In the external test set, the model with the highest increase in AUC was the *OP-Aug* model, with a gain of +1.45 in INbreast. This increase in AUC in the external test set showed that the models benefited from high-density synthetic data augmentation even when the training data is large and contains real BI-RADS D mammograms. The detailed metrics for the other data augmentation strategies can be found in Supplementary Material (Table D).

### 3.2.2 Low data availability scenario

The corresponding FROC curves of the detection models trained on the INbreast dataset are shown in Figure 7. All mass detection models were evaluated on the BI-RADS D test set from OPTIMAM Hologic. Table 4 summarizes the gains in AUC and the *p-values* for the best performing data augmentation strategies in INbreast. The detailed metrics for the other data augmentation strategies can be found in Supplementary Material (Table E).

The models trained using only synthetic BI-RADS D mammograms in training improved their performance in two out of four data augmentation strategies, namely, *OP-Aug* and *OP-CS-BC-Aug*. The *OP-Aug* model increased the AUC from 42.59% to 45.41%, a gain of +2.81 with a p-value smaller than 0.05 with respect to the baseline model. The OP-CS-BC-Aug model increased the baseline AUC by +1.05. When training the models with real and synthetic BI-RADS D mammograms, only the *OP-Aug* model obtained a gain in performance with respect to the baseline. However, the gain is higher than in other scenarios, increasing the AUC from 44.59% to 48.84%, a gain of +4.25.

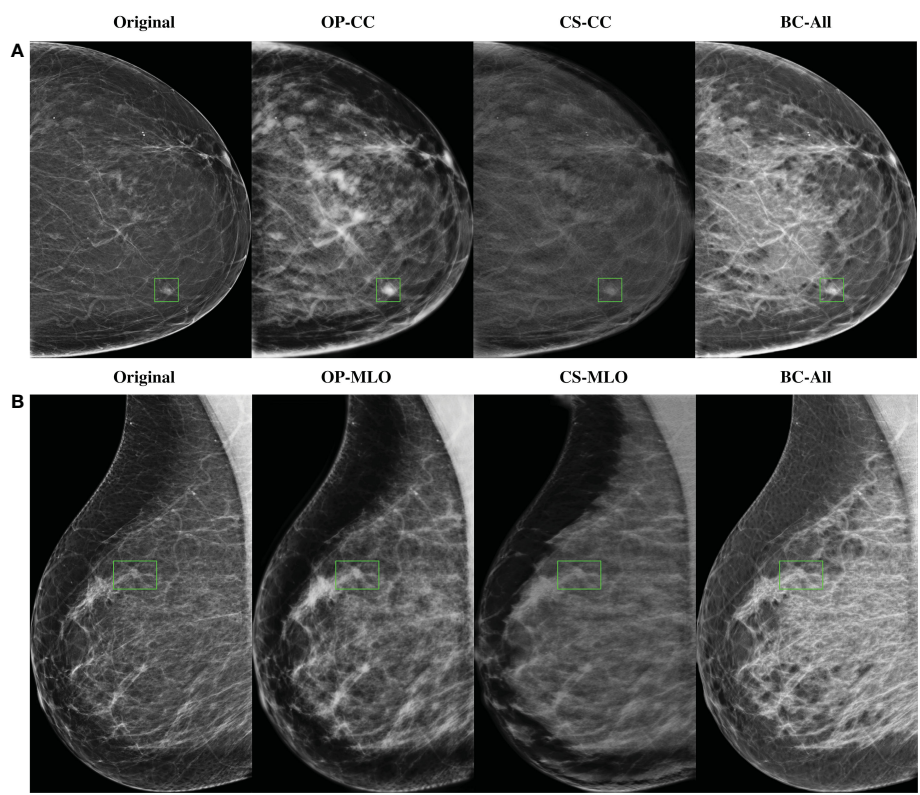

FIGURE 5
Samples of high-density synthetic mammograms generated with the different CycleGAN models. On the left, the input BI-RADS A mammogram from OPTIMAM Hologic dataset. **(A)** CC view, **(B)** MLO view.

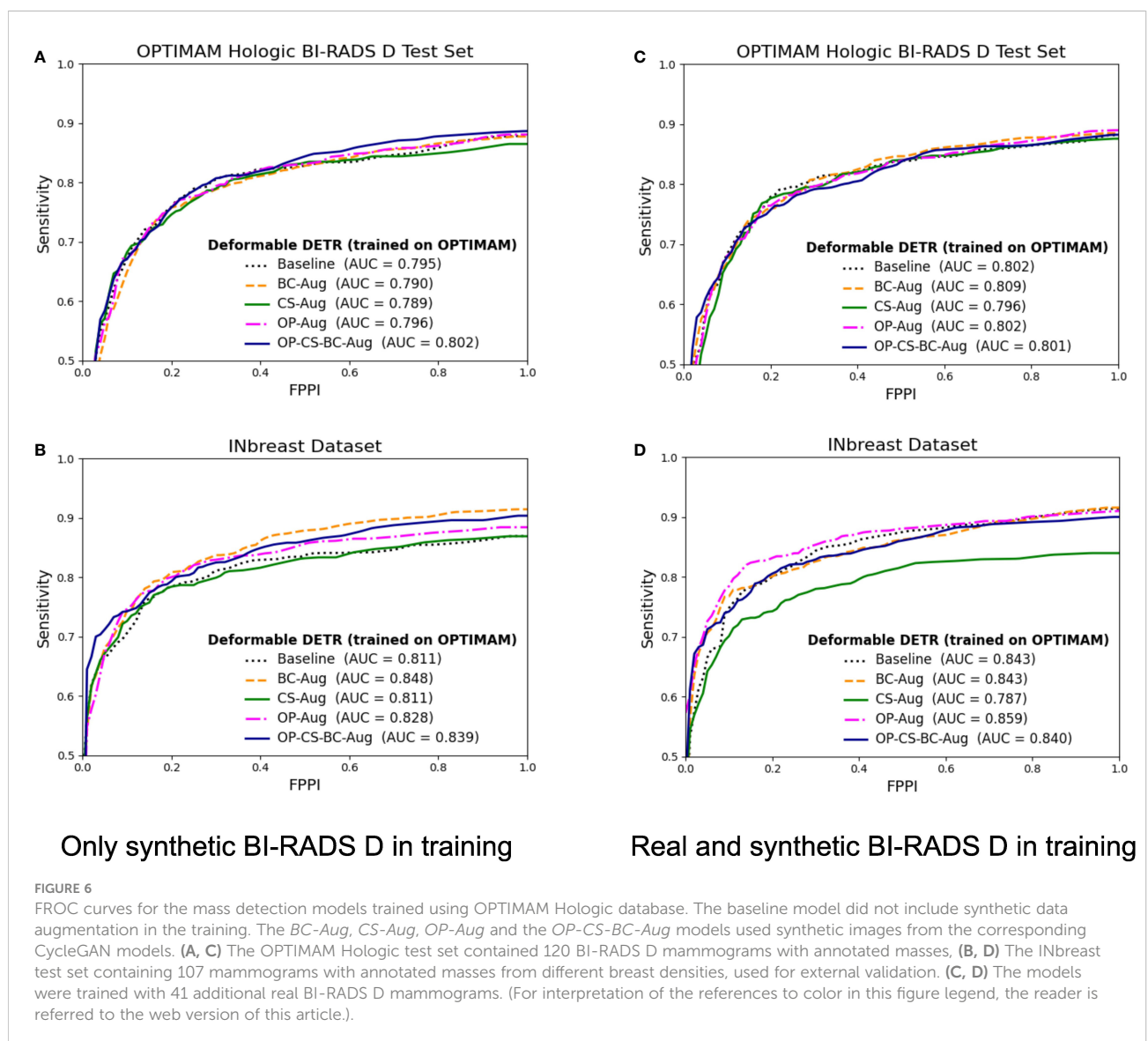

FIGURE 6
FROC curves for the mass detection models trained using OPTIMAM Hologic database. The baseline model did not include synthetic data augmentation in the training. The *BC-Aug*, *CS-Aug*, *OP-Aug* and the *OP-CS-BC-Aug* models used synthetic images from the corresponding CycleGAN models. **(A, C)** The OPTIMAM Hologic test set contained 120 BI-RADS D mammograms with annotated masses, **(B, D)** The INbreast test set containing 107 mammograms with annotated masses from different breast densities, used for external validation. **(C, D)** The models were trained with 41 additional real BI-RADS D mammograms. (For interpretation of the references to color in this figure legend, the reader is referred to the web version of this article.).

The AUC values in the OPTIMAM Hologic BI-RADS D test set are low in comparison with the AUCs in the large availability scenario. A first reason is that the INbreast models were trained with only 116 mammograms with masses. Second, the test set is from a different domain (OPTIMAM Hologic) due to the fact that the entire INbreast dataset was used as training set. The latter was motivated by the fact that the number of BIRADS D mammograms available in the INbreast dataset (Table 1) was deemed insufficient for representative performance evaluation.

## 3.3 Reader study outcomes

### 3.3.1 Receiver operating characteristic curves

The AUC of each reader ROC curve can be found in Table 5. On average, the synthetic mammograms of CC view from OPTIMAM CycleGAN were the most difficult to discriminate from original mammograms (0.615 AUC). Overall, the CC view looked more realistic to all readers than the MLO view. The BCDR model was the easiest for recognition of synthetic images, with an AUC of 0.824 in CC view and 0.954 in MLO view.

### 3.3.2 Qualitative analysis of synthetic FFDM

After completing the study, the readers evaluated the realism of the synthetic images and identified the common artifacts and failures.

TABLE 3 Performance values and statistical significance test results of the best data augmentation strategies for the models trained with OPTIMAM Hologic database.

| | | Only synthetic BI-RADS D in training | | | Synthetic and real BI-RADS D in training | | |
|---|---|---|---|---|---|---|---|
| | | FROC AUC | Gain | p-value | FROC AUC | Gain | p-value |
| OPTIMAM Hologic BI-RADS D Test Set | Baseline | 79.71% (78.44, 80.98) | Ref | Ref | 80.60% (79.20, 82.00) | Ref | Ref |
| | BC-Aug | 79.62% (77.83, 81.41) | -0.09 | 0.0064 | **81.10% (80.40, 81.80)** | **+0.50** | **0.2277** 0.2277 |
| | OP-Aug | 79.86% (78.30, 81.42) | +0.15 | 0.8269 | 80.75% (78.77, 82.73) | +0.15 | 0.5599 0.5599 |
| | OP-CS-BC-Aug | **80.95% (79.63, 82.27)** | **+1.24** | **0.0696** | 80.76% (79.92, 81.60) | +0.16 | 0.7921 0.7921 |
| INbreast Dataset (external validation) | Baseline | 81.51% (78.93, 84.09) | Ref | Ref | 84.71% (83.39, 86.03) | Ref | Ref |
| | BC-Aug | **85.66% (81.91, 89.41)** | **+4.15** | **0.0002** | 84.88% (82.86, 86.90) | +0.17 | 0.1666 0.1666 |
| | OP-Aug | 83.45% (80.03, 86.87) | +1.94 | 6.08e-05 | **86.16% (83.37, 88.95)** | **+1.45** | **0.0041** 0.0041 |
| | OP-CS-BC-Aug | 84.47% (82.32, 86.62) | +2.95 | 0.0008 | 84.29% (82.22, 86.36) | -0.42 | 0.0162 0.0162 |

The columns on the left correspond to the models trained without real BI-RADS D mammograms. The baseline models were trained without synthetic images. The 95% Confidence Intervals of the FROC AUC are in parenthesis. The *p-value* was computed using the DeLong method with a maximum of 10 FPPI. Bold values correspond to the best performing strategy. Ref corresponds to the reference method.

### 3.3.2.1 Common artifacts and failures

The most unrealistic features in order of importance were: 1) big concentrations of glandular tissue adjacent to pectoral muscle (Figure 8D); 2) dark small dots or ovals in the image like perforations (Figure 8A); 3) linear fragmentation of muscles; 4) distorted nipples (Figure 8C); and 5) lack of glandular tissue behind the nipple (Figure 8B). Some synthetic images missed to preserve the prepectoral fat or smooth contours at the interface with subcutaneous fat.

### 3.3.2.2 Realistic mammograms

Some synthetic high-density FFDMs that looked realistic to the radiologist are shown in Figure 9. The features that improved realism of synthetic images were 1) the presence of linear microcalcifications or roundish calcifications; 2) lymph nodes in the correct areas; 3) post-biopsy tissue markers; and 4) the correct distribution of dense tissue.

## 4 Discussion

The contributions in this study are two fold. First, synthesize high-resolution high-density FFDMs from different domains. Second, use the synthetic images as data augmentation to improve the mass detection performance in high-density breasts and multi-center datasets.

In our image synthesis experiments, we trained a total of five CycleGAN models using FFDMs from three different datasets to perform low-to-high density translation from real mammograms. Only healthy mammograms were used to train the image-to-image translation models. By training with healthy images, the models learned the healthy data distribution, hence, minimizing any risk of inserting hallucinated lesions into the synthetic images. As our goal was to use the synthetic mammograms to improve the mass detection, we confirmed that the CycleGAN models did not remove the masses from the original mammograms (see Figure 5).

To assess the quality of the synthetic images we calculated the commonly used FID metric between synthetic and real high-density mammograms. Since FID is not an absolute measure, we defined lower and upper bounds using real mammograms. The closer the synthetic images are to the real BI-RADS D images, the closer the FID should be to the lower bound. After evaluating the FID metric of the different CycleGAN models, we observed that the FID for the CC view was better than the one of the MLO view. The difficulty of synthesizing MLO view was previously mentioned in Korkinof et al. (24), most probably because MLO has greater complexity and more anatomical information than the CC view. However, the *OP-MLO* and *CS-MLO* models still had an FID score between the lower and upper bounds when the input images were from OPTIMAM Hologic.

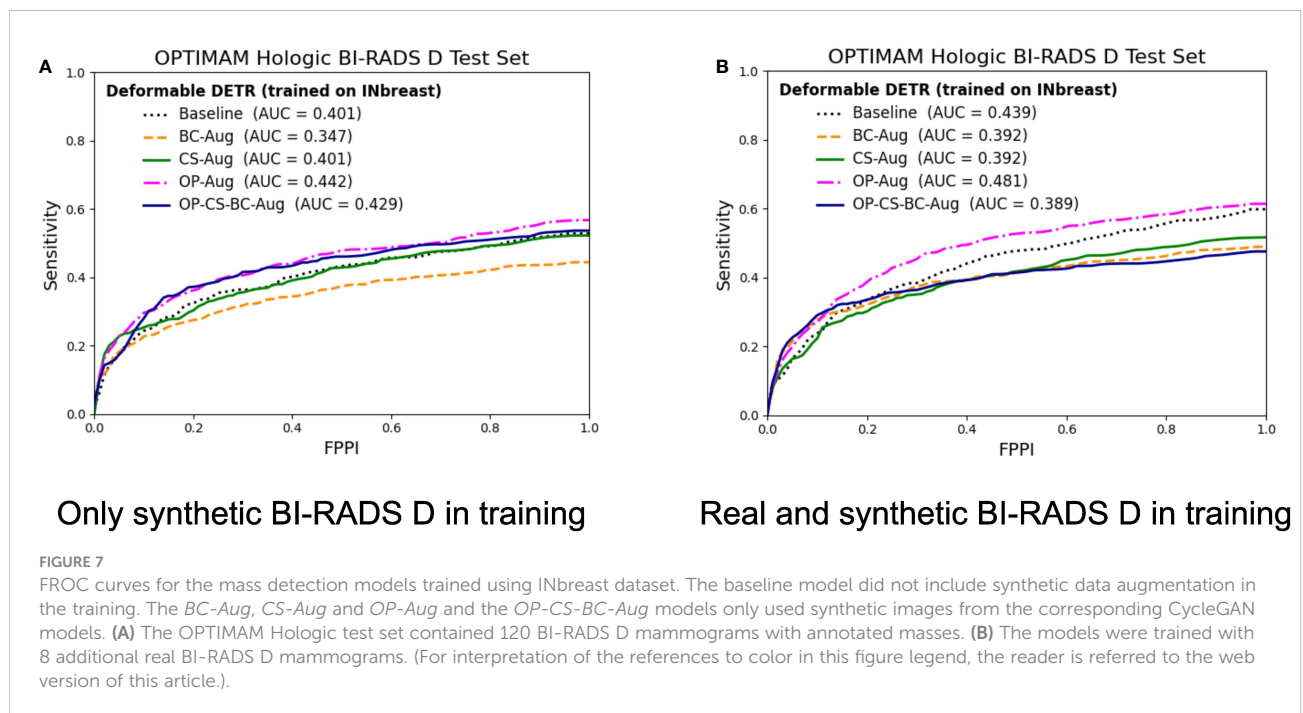

FIGURE 7
FROC curves for the mass detection models trained using INbreast dataset. The baseline model did not include synthetic data augmentation in the training. The *BC-Aug, CS-Aug* and *OP-Aug* and the *OP-CS-BC-Aug* models only used synthetic images from the corresponding CycleGAN models. **(A)** The OPTIMAM Hologic test set contained 120 BI-RADS D mammograms with annotated masses. **(B)** The models were trained with 8 additional real BI-RADS D mammograms. (For interpretation of the references to color in this figure legend, the reader is referred to the web version of this article.).

The *BC-All* CycleGAN had the largest FID score when the input images were from OPTIMAM Hologic, and the smallest FID when the input images were from INbreast. This is in line with the scanner manufacturer differences between both datasets. BCDR and INbreast were acquired with Siemens scanners. On the other hand, OPTIMAM and CSAW were acquired with Hologic Inc. scanners. Moreover, BCDR contains old digitized film mammograms while OPTIMAM is a Hologic digital mammography dataset with very specific image characteristics, i.e., it is very sharp, usually shows lymph nodes very well and it does visualize some skin of the breast.

To further evaluate the clinical realism of the synthetic images, we performed a reader study involving two breast radiologists and one surgical oncologist. When the CycleGAN models trained using OPTIMAM Hologic and CSAW datasets (*OP-CC, OP-MLO, CS-CC, CS-MLO*) were used to insert density onto low density mammograms from OPTIMAM Hologic, it was much more difficult for the readers to differentiate between original and synthetic mammograms. In that case, the readers had to look for anatomical disparities and inadequacies to spot the difference. Both OPTIMAM and CSAW FFDMs were acquired with an Hologic scanner. On the other hand, all the readers could easily identify the synthetic images generated with the *BC-All* model. As previously mentioned, BCDR contains old digitized film mammograms acquired with a Siemens scanner. Considering that, we can conclude that domain disparities between the acquisition settings of the source and the target domains have a big impact on the perceptual realism of the

TABLE 4 Performance values and statistical significance of the best data augmentation strategies for the models trained with INbreast dataset.

| | | Only synthetic BI-RADS D in training | | | Synthetic and real BI-RADS D in training | | |
|---|---|---|---|---|---|---|---|
| | | FROC AUC | Gain | *p-value* | FROC AUC | Gain | *p-value* |
| OPTIMAM Hologic BI-RADS D Test Set (external validation) | Baseline | 42.59% (39.73, 45.45) | Ref | Ref | 44.59% (42.87, 46.31) | Ref | Ref |
| | OP-Aug | **45.41% (42.70, 48.12)** | **+2.81** | **4.58e-22** | **48.84% (46.21, 51.47)** | **+4.25** | **1.43e-27** |
| | OP-CS-BC-Aug | 43.65% (40.64,46.66) | +1.05 | 2.32e-11 | 39.74% (35.72, 43.76) | -4.85 | 9.40e-10 |

The columns on the left correspond to the models trained without real BI-RADS D mammograms. The baseline models were trained without synthetic images. The 95% Confidence Intervals of the FROC AUC are in parenthesis. The *p-value* was computed using the DeLong's method with a maximum of 10 FPPI. Bold values correspond to the best performing strategy. Ref corresponds to the reference method.

TABLE 5 Reader test results: Area Under the Curve (AUC) from the Receiver Operating Characteristics (ROC) curve of each CycleGAN model and view (CC, MLO).

| | OPTIMAM | | CSAW | | BCDR | |
|---|---|---|---|---|---|---|
| | CC | MLO | CC | MLO | CC | MLO |
| Reader A | 0.580 | 0.664 | 0.651 | 0.718 | 0.818 | 0.962 |
| Reader B | 0.576 | 0.893 | 0.887 | 0.758 | 0.871 | 0.956 |
| Reader C | 0.689 | 0.673 | 0.489 | 0.636 | 0.784 | 0.944 |
| Average ± std | 0.615 ± 0.052 | 0.743 ± 0.105 | 0.675 ± 0.163 | 0.704 ± 0.050 | 0.824 ± 0.037 | 0.954 ± 0.007 |

Reader A: 9+ years of experience as a breast radiologist. Reader B: 7+ years of experience as a breast radiologist. Reader C: surgical oncologist with +12 years of experience in image guided breast biopsy and lesion localization techniques.

synthetic images. In addition, the synthetic mammograms from the MLO view were easier to identify than the ones from the CC view and this observation correlates with the higher FID scores of the synthetic images from the MLO CycleGAN models (*OP-MLO* and *CS-MLO*).

In our mass detection experiments, different data augmentation strategies were tested to improve the mass detection of the baseline model. First, three mass detection models were trained using synthetic images from the different CycleGANs with a proportion of 1:1 – 1 real and 1 synthetic. Second, a fourth mass detection model was trained with synthetic images from all the CycleGAN models, the *OP-CS-BC-Aug*, with a proportion of 1:3 – 1 real and 3 synthetic. To evaluate the effect of the different data augmentation strategies under diverse training conditions, we defined four training scenarios for the mass detection models. Two scenarios involving the amount of data available for training, and the other two involving the exclusion or inclusion of the real BI-

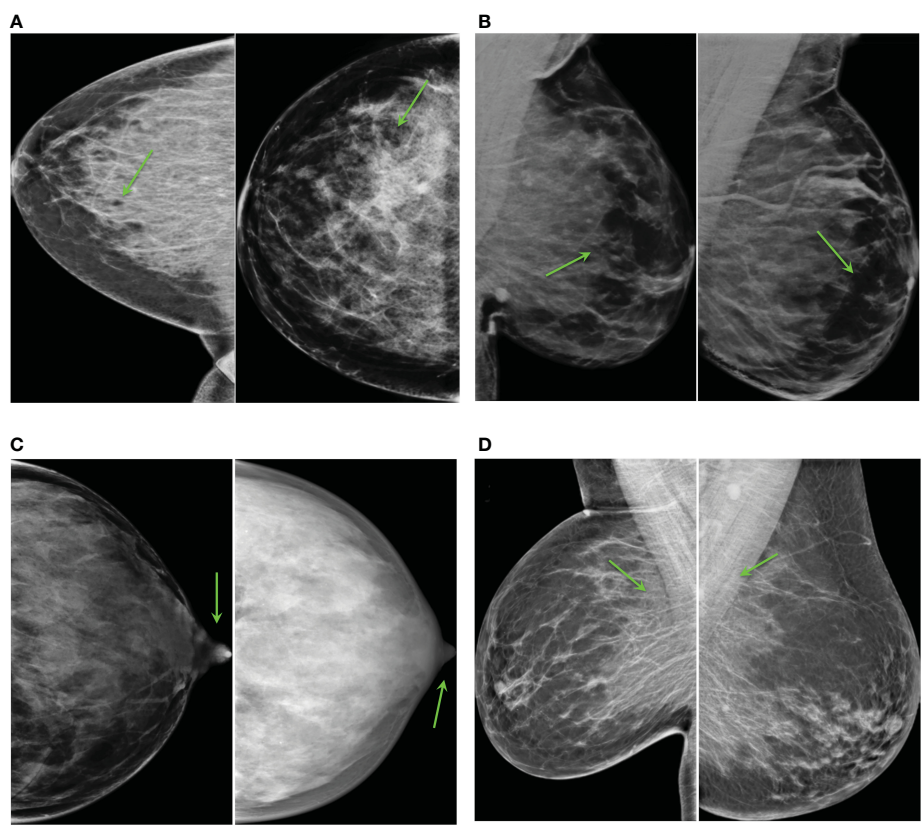

FIGURE 8
Common artifacts of synthetic high-density mammograms, (A) dark dots or ovals, (B) lack of glandular tissue behind the nipple, (C) distorted nipples, (D) big concentrations of glandular tissue adjacent to the pectoral muscle.

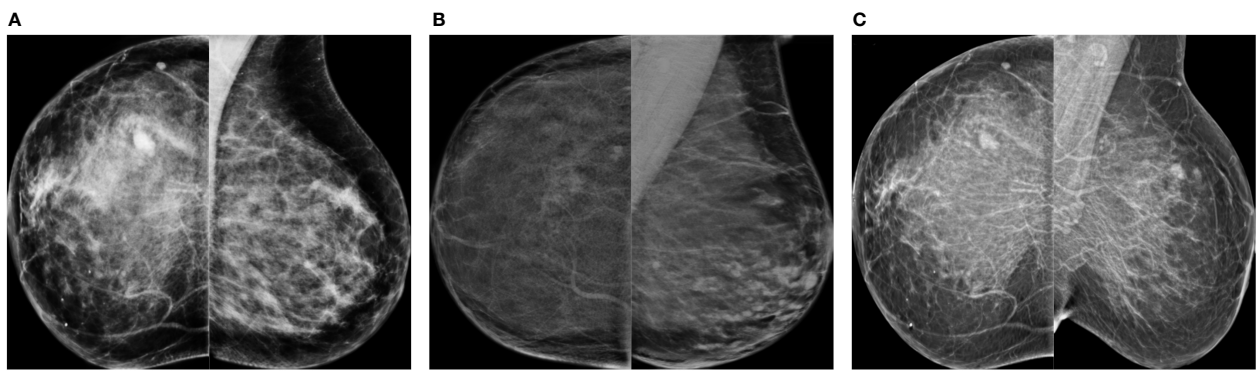

FIGURE 9
Synthetic high-density mammograms that looked realistic to the radiologist. Samples from the different CycleGAN models: **(A)** *OP-CC* and *OP-MLO*, **(B)** *CS-CC* and *CS-MLO*, **(C)** *BC-All*.

RADS D mammograms in training. By excluding the real BI-RADS D mammograms, we tested if the synthetic high-density mammograms could replace the real BI-RADS D images, simulating a plausible clinical scenario in which no BI-RADS D images were available for training. In the large availability scenario, the data augmentation strategies did not improve the baseline performance with statistically significant difference. However, the synthetic data helped the models to generalize better and increased the performance in the external dataset (INbreast). When training only with synthetic BI-RADS D images, the *BC-Aug* model improved the AUC by 4.15% in INbreast. When the real BI-RADS D images were present in training, the *OP-Aug* model improved the AUC by 1.45%. In the low data availability scenario, the *OP-Aug* data augmentation strategy improved the AUC significantly for both scenarios in the out-of-domain test set.

As we hypothesized, the CycleGAN models not only learned how to translate from low-to-high density but also preserved the domain characteristics from their respective training datasets during translation. The domain characteristics comprise the differences in image quality, acquisition settings and scanner manufacturers.

Based on our experimental results, we did not observe a consistent association between the FID scores and the success of the corresponding data augmentation in mass detection. For instance, in the low-data availability scenario, the CycleGANs trained on OPTIMAM Hologic had a comparably high FID but the *OP-Aug* detection model obtained the highest increase in the AUC. Moreover, the *OP-Aug* detection model yielded a consistent improvement in the four training scenarios. The *OP-CC* and *OP-MLO* CycleGANs were trained with a larger number of training images than the CSAW and BCDR CycleGANs. Overall, the number of images used to train the CycleGAN models seemed to have a higher importance for the detection performance gain compared to the FID score.

One of the limitations of our work involves the availability of healthy BI-RADS D mammograms to train the generative models. From the results, the data augmentation that obtained more consistent gains under all four scenarios was the *OP-Aug*, using CycleGANs trained with the OPTIMAM Hologic dataset. As shown in Figure 3, *OP-MLO* and *OP-CC* were trained with more images than the other CycleGANs, showing the big impact of data for training more robust CycleGANs. In this regard, fairer comparisons among generative models could be achieved if more images from the CSAW and BCDR datasets would be available.

Data scarcity is one of the most common and general limitations in medical imaging AI research. In tasks where patch-based approaches are sufficient, extracting multiple samples from one scan can aid to overcome the data scarcity issue (47). However, this limitation is further exacerbated in tasks where a single subject scan can only be used as a single sample in the training of deep learning methods. Moreover, the high resolution nature of the breast FFDMs makes it even more challenging. We believe that this study is not only an important step towards mitigating data scarcity and class imbalance, but also demonstrates the importance of fair AI in clinical practice. However, there is still room for improvement to increase the fairness of AI models for women with high-density breasts in mammography screening.

Future research may focus on evaluating the potential of the high-density synthetic mammograms in other downstream tasks. We analyzed the high-density mammogram synthesis *via* the downstream task of breast mass detection. However, there is a multitude of further applications of our proposed low-to-high breast density translation. For instance, our method can be applied to breast mass segmentation or tumor malignancy classification. Furthermore, our synthetic images can expand the radiologist curricular and professional training programs. For instance, for training purposes radiologist candidates would need to detect and accurately annotate the same lesion in both high and low-density breast, which allows to measure inter- and intra-observer variability per lesion and breast density level. This way our method provides flexibility to personalize and adjust the radiologist training to specific scenarios (e.g. very high density breasts with very small-sized tumor lesions).

## 5 Conclusion

In this study, we evaluated different CycleGAN models for high-density FFDM synthesis from three different datasets and acquisition pipelines, comprising two scanner manufacturers – i.e. Hologic and Siemens. Moreover, we applied different synthetic data augmentation strategies to improve the mass detection performance of a deep-learning based model. Even though the improvements were not always statistically significant for models trained in the large data availability scenario, the results demonstrated that the data augmentation helped to improve the mass detection in out-of-domain datasets, improving the domain generalization of the final model. The models trained in the low data availability scenario obtained more benefit from the data augmentation, with a maximum gain of 4.25% in AUC for the model trained with synthetic images generated with the OPTIMAM Hologic CycleGANs. Finally, a reader study involving three expert radiologists evaluated the perceptual realism of the synthetic mammograms, concluding that the quality of CC view synthetic images is higher than the mammograms from MLO view. Our study is the first one to synthesize high-resolution FFDMs with increased density and showed the potential of including the generated images in the data augmentation pipeline to improve the generalization and performance of downstream tasks using mammography images, such as mass detection.

## Data availability statement

The data analyzed in this study is subject to the following licenses/restrictions: A subset of the OPTIMAM Database was obtained as part of the data-sharing agreement with the University of Barcelona in 2021. Requests to access these datasets should be directed to Cancer Research Horizons, horizons@cancer.org.uk. The CycleGAN models and weights mentioned in this study can be found in the medigan library (https://github.com/RichardObi/medigan).

## Ethics statement

Written informed consent for participation was not required for this study in accordance with the national legislation and the institutional requirements.

## Author contributions

LG, KK, RO, and OD contributed to the conception and design of the study. LG organized the datasets, performed the experiments and designed the reader study. MB, AC, and JR participated in the reader study and clinical assessment of the results. FS contributed with one dataset (CSAW) in this study. LI and KL supervised the research. All authors contributed to article revision and read and approved the submitted version.

## Funding

This project has received funding from the European Union's Horizon 2020 research and innovation programme under grant agreement No 952103. This work was supported in part by the MICINN Grant RTI2018-095232-B-C21 and 2017 SGR 1742.

## Acknowledgments

A subset of the OPTIMAM database was obtained as part of the data-sharing agreement with the University of Barcelona in 2021. Special thanks to Dr Melissa Hill (Volpara Health) for sharing the breast density information of the OPTIMAM subset and to Dr Premkumar Elanvogan for sharing the base code to prepare the ImageJ plugin for the reader study.

## Conflict of interest

The authors declare that the research was conducted in the absence of any commercial or financial relationships that could be construed as a potential conflict of interest.

## Publisher's note

All claims expressed in this article are solely those of the authors and do not necessarily represent those of their affiliated organizations, or those of the publisher, the editors and the reviewers. Any product that may be evaluated in this article, or claim that may be made by its manufacturer, is not guaranteed or endorsed by the publisher.

## Supplementary material

The Supplementary Material for this article can be found online at: https://www.frontiersin.org/articles/10.3389/fonc.2022.1044496/full#supplementary-material